\documentclass{ametsocV6.1}
\providecommand{\appendix}{\par\setcounter{section}{0}\setcounter{subsection}{0}\renewcommand{\thesection}{\Alph{section}}}

\usepackage{amsmath,amsfonts,amssymb,bm}
\usepackage{mathptmx}
\nolinenumbers

\usepackage[T1]{fontenc}
\usepackage{color,soul}
\usepackage{mdframed}
\usepackage{xcolor}

\usepackage{textcomp}
\usepackage{nicefrac} 
\usepackage{xfrac}
\usepackage{mathtools}
\usepackage{newtxmath}
\usepackage{newtxtext}

\title{Bayesian Deep Learning for Convective Initiation Nowcasting Uncertainty Estimation}
\begin{document}
\authors{Da Fan$^a$\correspondingauthor{Da Fan, dxf424@psu.edu}, David John Gagne II$^b$, Steven J. Greybush$^{a,c}$, Eugene E. Clothiaux$^a$, John S. Schreck$^b$, and Chaopeng Shen$^d$}\
\affiliation{\aff{a}Department of Meteorology and Atmospheric Science, The Pennsylvania State University, University Park, Pennsylvania\\
\aff{b}National Center for Atmospheric Research, Boulder, Colorado\\\aff{c}Institute for Computational and Data Sciences, The Pennsylvania State University, University Park, Pennsylvania\\\aff{d}Department of Civil and Environmental Engineering, The Pennsylvania State University, University Park, Pennsylvania}
%


\abstract
{This study evaluated the probability and uncertainty forecasts of five recently proposed Bayesian deep learning methods relative to a deterministic residual neural network (ResNet) baseline for 0-1 h convective initiation (CI) nowcasting using GOES-16 satellite infrared observations. Uncertainty was assessed by how well probabilistic forecasts were calibrated and how well uncertainty separated forecasts with large and small errors. Most of the Bayesian deep learning methods produced probabilistic forecasts that outperformed the deterministic ResNet, with one, the initial-weights ensemble + Monte Carlo (MC) dropout, an ensemble of deterministic ResNets with different initial weights to start training and dropout activated during inference, producing the most skillful and well-calibrated forecasts. The initial-weights ensemble + MC dropout benefited from generating multiple solutions that more thoroughly sampled the hypothesis space. The Bayesian ResNet ensemble was the only one that performed worse than the deterministic ResNet at longer lead times, likely due to the challenge of optimizing a larger number of parameters. To address this issue, the Bayesian-MOPED (MOdel Priors with Empirical Bayes using Deep neural network) ResNet ensemble was adopted, and it enhanced forecast skill by constraining the hypothesis search near the deterministic ResNet hypothesis. All Bayesian methods demonstrated well-calibrated uncertainty and effectively separated cases with large and small errors. In case studies, the initial-weights ensemble + MC dropout demonstrated better forecast skill than the Bayesian-MOPED ensemble and the deterministic ResNet on selected CI events in clear-sky regions. However, the initial-weights ensemble + MC dropout exhibited poorer generalization in clear-sky and anvil cloud regions without CI occurrence compared to the deterministic ResNet and Bayesian-MOPED ensemble.
}

\maketitle
\section{Introduction}
Accurate prediction of convective initiation (CI) remains difficult for both empirical and numerical weather prediction (NWP) models \citep[e.g.,][]{Mecikalski2015, Lawson2018, Cintineo2020a} due to its sensitivity to sub-kilometer processes. Our inability to provide timely and precise CI forecasts often results in delayed responses to convective hazards such as hailstorms and tornadoes \citep[][]{Brooks2008, Dixon2011}. Recently, several novel machine learning (ML) models \citep[][]{Lee2017, Sun2023, Fan2024} have been developed to enhance short-term CI forecast skill. \cite{Lee2017} developed a random forest model to predict CI for tracked cloud objects using cloud characteristics from satellite observations. Using similar features, \cite{Fan2024} applied a convolutional neural network that produced skillful CI forecasts at lead times up to 1 hour. \cite{Sun2023} developed a convolutional recurrent neural network that leverages spatiotemporal features from both satellite and radar data, demonstrating good performance in predicting several CI cases at lead times up to 30 min. However, due to the lack of very fine-resolution observations, these models are unable to resolve characteristics on scales of O(1 km), such as differential topography and fast-growing cumulus clouds, and still generate incorrect CI forecasts in some scenarios. An assessment of uncertainty is thus needed to arrive at a reasonable confidence level in any prediction and to provide a better understanding of model behavior. 

Robust estimates of predictive uncertainty and well-calibrated forecasts strengthen the reliability of forecasts and aid in the decision-making process \citep[e.g.,][]{Nadav-Greenberg2009, Kendall2017}. Uncertainty estimates support identification of inherently challenging cases and indicate when the model is operating beyond the scope of the training data \citep{Kendall2017, Karpatne2017, Fang2020}. Uncertainty decomposes into aleatoric, due to internal variability within the training data, and epistemic, arising from limitations in model architecture and the constrained realm of training data, uncertainties \citep{Kendall2017}. Recently, uncertainty quantification of ML models for weather and climate applications has received much attention \citep{McGovern2017,Haynes2023,Schreck2023}. Popular approaches include parametric probabilistic models \citep{Ghazvinian2021, Rasp2018, Chapman2022, Guillaumin2021, Barnes2021, Foster2021, Delaunay2022, Gordon2022}, quantile-based distribution models \citep{Scheuerer2020, Bremnes2020, Yu2020, Schulz2022}, evidential models \citep{Sensoy2018, Amini2020, Ulmer2023, Schreck2023}, and Bayesian model averaging \citep[BMA;][]{Raftery2005}. BMA was originally introduced into the meteorological community \citep{Raftery2005} to calibrate forecast ensembles by combining forecasts from multiple models and analyses. More recently, it has been applied in the ML community \citep{Wilson2020} to represent predictive distributions generated by aggregating forecasts sampled from Bayesian models.

Parametric probabilistic models predict the parameters of a probability distribution, such as a Gaussian distribution, but they only account for aleatoric uncertainty \citep{Nix1994}. Evidential models modify the output layer of a single deterministic model to estimate the parameters of a higher-order evidential distribution to capture both aleatoric and epistemic uncertainty \citep{Gelman2003}, but might underrepresent epistemic uncertainty. BMA generates the predictive distribution by averaging the predictions of different models, weighted by their posterior probabilities, and estimates uncertainty from the predictive distribution. BMA is able to accurately estimate both aleatoric and epistemic uncertainty. As one BMA approach, Bayesian neural networks \citep[BNNs;][]{neal2012bayesian,Ortiz2023} learn the distribution of each parameter and provide a robust estimate of the predictive distribution by approximating the posterior distribution. \cite{Ortiz2022} and \cite{Ortiz2023} applied BNNs to classify precipitation type and generate synthetic microwave images from satellite infrared observations, respectively, achieving performances comparable to a deterministic model while providing robust uncertainty estimates. However, BNNs require more weights than deterministic models with the same architecture and often struggle to converge to a solution that performs comparably to deterministic models in complex applications \citep{Krishnan2020}. To address the convergence issue, the MOdel Priors with Empirical Bayes using Deep neural network (MOPED) method was introduced by \cite{Krishnan2020}. This method initializes the weight priors in BNNs using pretrained deterministic models, accelerating training convergence and enhancing performance for different tasks \citep{Krishnan2020, Zhang2022,Milanes-Hermosilla2023}. The informed priors in the MOPED method were also argued to improve generalization by BNNs \citep{Zhang2022}. Additionally, applying Monte Carlo dropouts to randomly deactivate neurons of deterministic models can be interpreted as approximating the posterior with a set of sampled points \citep{Gal2016}. 

The initial-weights ensemble approach, often referred to as the "deep ensemble" \citep{Lakshminarayanan2017}, provides accurate and well-calibrated predictive distributions. An initial-weights ensemble comprises an ensemble of deterministic models, each trained with a different set of random initial weights. By searching for various solutions in the hypothesis space, the initial-weights ensemble offers a better approximation of the true predictive distribution than BNNs \citep{Ovadia2019,Wilson2020}. \cite{Filos2019} demonstrated that initial-weights ensemble methods generate better forecasts than deterministic models and provide accurate uncertainty estimates. \cite{Wilson2020} suggested that functional diversity is important for a good approximation of the predictive distribution.

In this study, we extend the work of \cite{Fan2024} to explore CI nowcasting skill and uncertainty estimates produced by Bayesian deep learning methods, including Bayesian and initial-weights neural network ensembles. The objectives of our study include: (1) systematically comparing forecast skill of Bayesian deep learning methods for CI nowcasting; (2) evaluating the relationship between performance and predictive uncertainty; and (3) investigating convergence issues in a Bayesian neural network and the impact of the MOPED method when incorporated within it. The rest of this paper is organized as follows. Section 2 describes CI identification and data preprocessing. Section 3 lays out the model architectures and evaluation methods used in the study. Section 4 evaluates the performances of the Bayesian deep learning methods and their uncertainties. Section 5 discusses the convergence issues of Bayesian neural networks. Finally, Section 6 presents the main findings and limitations of the study, and includes concluding remarks.

\section{Data}

\begin{figure*}[!ht]
\centerline{\includegraphics[width=\textwidth,angle=0]{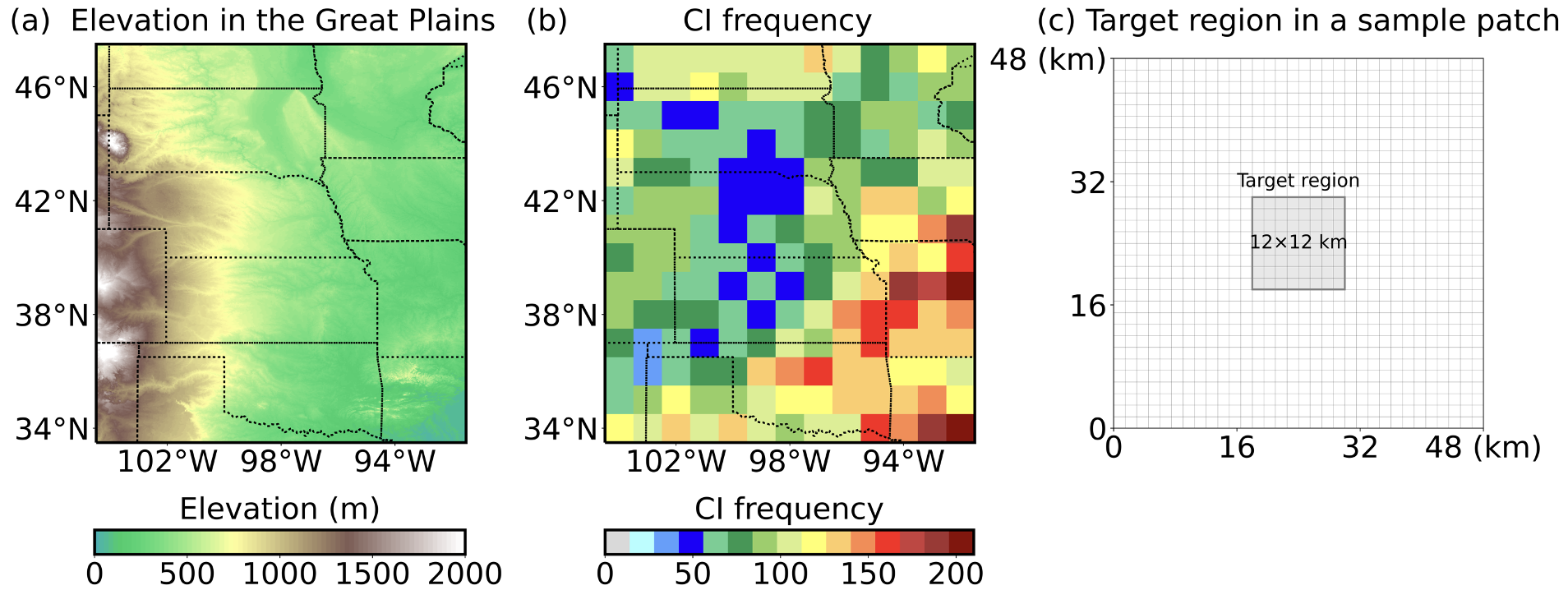}}
\caption{(a) The Great Plains region upon which our study focused along with its topography (colors). (b) The frequency of CI events on a 1° grid for June 2021. (c) The 12 km ×12 km target region in a sample patch for the generalization test.}\label{fig1}
\end{figure*}

The dataset in this study is the same as that in \cite{Fan2024}. We focus on CI events obtained from the Multi-Radar Multi-Sensor dataset \citep[MRMS;][]{Lakshmanan2006, Lakshmanan2007} across the Great Plains region of the United States (Fig. \ref{fig1}a). The predictors are infrared observations from the GOES-16 Advanced Baseline Imager (ABI). More details are below.

\subsection{CI identification}

CI clusters are identified using MRMS composite reflectivity on a 0.01° spatial grid and a 2-min temporal grid. Pixels are defined as being part of a CI cluster when the following conditions \citep[adapted from][]{Colbert2019} are met:

(1) composite reflectivity $\geq$ 35 dBZ,

(2) in the preceding 11 min, no points within 15 km exhibit composite reflectivity $\geq$ 35 dBZ, and

(3) in the preceding 30 min, no points within 5 km exhibit composite reflectivity $\geq$ 35 dBZ.

\noindent The first condition identifies pixels associated with convection and the subsequent two eliminate pixels with preexisting convection. Storm tracking and CI identification are done through the w2segmotionll algorithm \citep{Lakshmanan2010} and a modified best-track algorithm \Citep{Lakshmanan2015}. For a more detailed explanation see \cite{Fan2024}. 

\subsection{Data preprocessing}
We use an object-based method to identify localized environments within which to predict CI. CI events are 48-km by 48-km square patches centered on a CI cluster. Most (approximately 91\%) of the non-CI events are patches of the same size that are nearest neighbors to CI events and contain no CI cluster of their own. The remaining (approximately 9\%) non-CI events are same size patches randomly extracted across the Great Plains area and neither contain a CI cluster nor are a nearest neighbor to a patch that does. To avoid the impacts of class imbalance on model performance, non-CI events, the majority class, are undersampled to produce a balanced dataset, including 58\% CI and 42\% non-CI events. Binary labels are used to classify CI and non-CI events. The entire dataset consists of 94,618 samples. Of these, 45,077 and 19,320 samples from June and July 2020 are used for training and validation, respectively, while 30,221 samples collected in June 2021 are for testing. Fig. \ref{fig1}b shows CI frequency across the Great Plains in June 2021. The regions in eastern Kansas, central and eastern Oklahoma, Missouri, and Arkansas exhibit more frequent CI than the other regions.

Predictors for CI and non-CI events are brightness temperatures \citep[BTs; Table 1 in][]{Fan2024} within the 48-km by 48-km square patches extracted from seven GOES-16 ABI infrared channels with approximately 2-km native horizontal resolution and 5-minute temporal sampling. Predictors are obtained at lead times ranging from 60 minutes down to 10 minutes before the occurrence time of an event. For each lead time, models are trained using the predictors specific to that lead time and predictands. During data preprocessing, parallax error, the horizontal spatial displacement of clouds when projected to the reference surface along the view of the satellite, is corrected using the cloud-top height (ACHA) product of GOES-16, following \cite{Zhang2019}. Then, GOES-16 ABI BTs are remapped from their native approximately 2-km mesh to a 1.5-km uniform mesh through linear interpolation, generating 32×32 input BT values to facilitate development of a deep neural network model. See \cite{Fan2024} for complete details.


\section{Methods}
This section introduces the deterministic model, Bayesian deep learning models, and model evaluation. Figure \ref{fig2} depicts the architecture of the deterministic and Bayesian deep learning models, whereas Figure \ref{fig3} presents their conceptual diagram to show their strategies for hypothesis exploration within the function space $\mathcal{F}$.

\subsection{Deterministic ResNet}
\begin{figure*}[!ht]
\centerline{\includegraphics[width=\textwidth,angle=0]{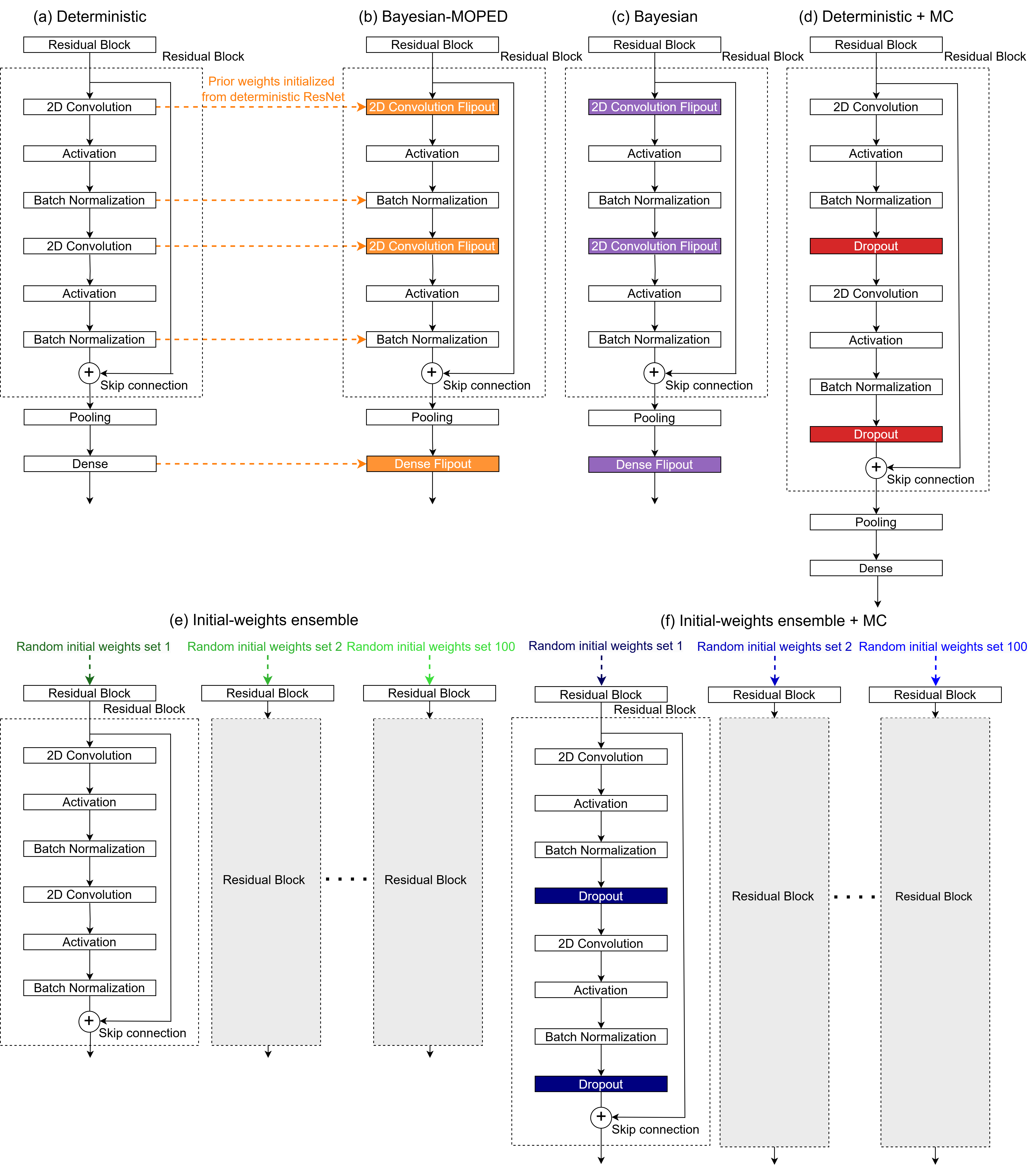}}
\caption{A depiction of the ResNet model and five Bayesian deep learning methods based on it. Inside each dashed rectangle is a single residual block. The changes from the ResNet model to the five Bayesian deep learning methods are highlighted in different colors. The orange dashed arrows indicate that weight priors in the Bayesian model with MOPED are initialized by the pretrained weights in the deterministic model.}\label{fig2}
\end{figure*}

The residual neural network (ResNet) architecture with 6 residual blocks and 15 convolutional layers (Fig. \ref{fig2}a) showed skillful CI forecasts in \cite{Fan2024} and is used here as the baseline model. The strength of the ResNet is attributed to the skip connection within each residual block (Fig. \ref{fig2}a). The skip connection allows the flow of information from earlier layers directly into later layers, mitigating the gradient vanishing problem prevalent in deep neural networks. The skip connection enabled a significantly deeper neural network than possible without it and possible with a traditional neural network. In \cite{Fan2024}, the ResNet extracted spatial features of clouds and moisture to enhance its nowcast skills. The ResNet was trained using the binary cross-entropy loss and optimized on the area-under-curve (AUC) score using Earth Computing Hyperparameter Optimization (ECHO: \url{https://doi.org/10.5281/zenodo.7787022}). For full details on training and optimization see \cite{Fan2024}. The ResNet was built using the Keras library \citep{chollet2015keras} with a Tensorflow low-level backend \citep{Abadi2016}.

\clearpage

In this study, the ResNet generates forecasts under fixed weights with dropout deactivated in the inference stage, and is herein referred to as the ``deterministic ResNet''. Within the hypothesis space $\mathcal{H} \subseteq \mathcal{F}$, the deterministic ResNet (Fig. \ref{fig3}a) learns a single hypothesis $h_{d}$ from the training data. The discrepancy between $h_{d}$ and the ground truth function $f^{*}$ is attributable to several factors: the stochastic dependency between the inputs $x$ and outputs $y$, constraints on the hypothesis space placed by the model architecture, and the amount and quality of training data. In the following section, we will introduce the modifications that transform the deterministic ResNet into the Bayesian methods.

\subsection{Bayesian model averaging (BMA)}

Unlike the deterministic ResNet which produces a single prediction, BMA provides a distribution of predictive probability using the mean forecasts of different models, weighted by their posterior probabilities. Formally, we assume a training dataset $D = \left\{  \mathbf{X}, \mathbf{Y} \right\}$ with inputs $\mathbf{X} = \{x_1, x_2 \ldots, x_n\}$ and their labels $\mathbf{Y} = \{y_1, y_2 \ldots, y_n\}$, where $n$ represents the dataset size, $x_n$ represents an input vector for a sample, and $y_n$ represents the corresponding label. For new inputs $\mathbf{X}^{*}$, the predictive distribution $p(\mathbf{Y}^* | \mathbf{X}^*)$ is given by
\begin{equation}
\label{eq:1}
p(\mathbf{Y}^* | \mathbf{X}^*) = \int p(\mathbf{Y}^* | \mathbf{X}^*, \mathbf{W})p(\mathbf{W} | \mathbf{X}, \mathbf{Y})d\mathbf{W},
\end{equation}
where $p(\mathbf{Y}^*|\mathbf{X}^*, \mathbf{W})$ is the forecast probability of the model with the weight $\mathbf{W}$, and $p(\mathbf{W} | \mathbf{X}, \mathbf{Y})$ is the posterior distribution of the model weight $\mathbf{W}$, updated based on the observed training data. BNNs estimate the predictive distribution through approximating the posterior distribution for $\mathbf{W}$, whereas the initial-weights ensemble approximates the predictive distribution directly through exploring multiple solutions in the hypothesis space.

\subsubsection{BAYESIAN RESNET}

\begin{figure*}[h]
\centerline{\includegraphics[width=\textwidth,angle=0]{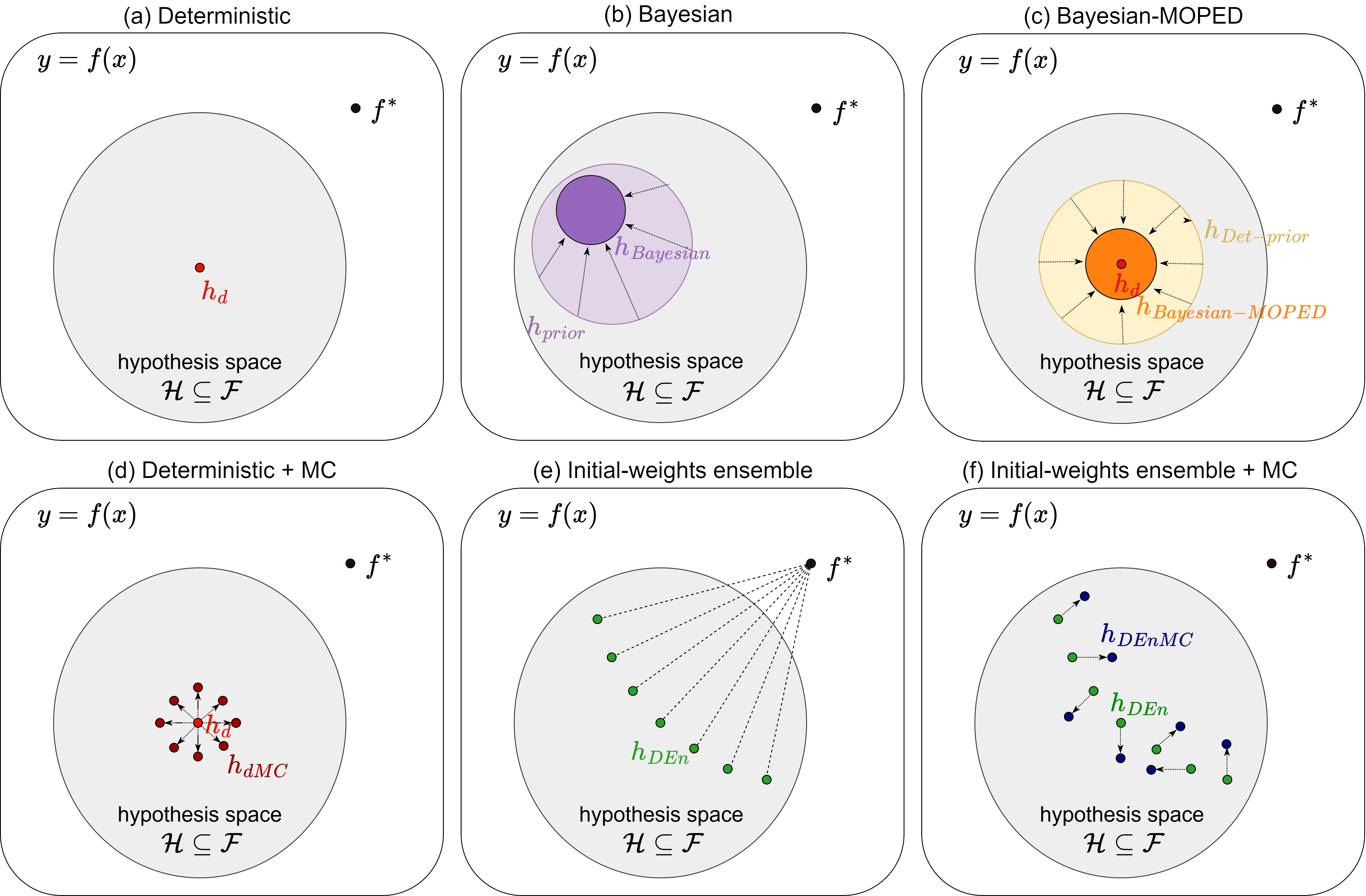}}
\caption{Conceptual diagrams illustrating hypothesis search strategies for the (a) deterministic ResNet, (b) Bayesian ResNet, (c) Bayesian-MOPED ResNet, (d) deterministic ResNet + MC dropout, (e) initial-weights ensemble, and (f) initial-weights ensemble + MC dropout methods within the entire function space $\mathcal{F}$. $\mathcal{H}$ is the hypothesis space. $x$ and $y$ are the inputs and outputs, respectively. $f$ is the function mapping $x$ to $y$. $f^{*}$ is the ground truth function. $h_{d}$, $h_{dMC}$, $h_{DEn}$, and $h_{DEnMC}$ are the point hypothesis generated by the deterministic ResNet, deterministic ResNet + MC dropout, initial-weights ensemble of ResNets, and initial-weights ensemble of ResNets + MC dropout, respectively. $h_{prior}$ represents a randomly chosen hypothesis distribution for the Bayesian ResNet, whereas $h_{Det-prior}$ indicates an informed prior distribution of hypotheses near the deterministic ResNet hypothesis $h_{d}$ for the Bayesian-MOPED ResNet. Within the prior space, $h_{Bayesian}$ and $h_{Bayesian-MOPED}$ represent the posterior distribution over the hypothesis space generated by the Bayesian and Bayesian-MOPED ResNets, respectively. Small arrows indicate the connections between hypotheses.}\label{fig3}
\end{figure*}

Bayesian neural networks \citep[BNNs;][]{neal2012bayesian} provide a probabilistic interpretation of neural networks by treating their weights as random variables characterized by probability distributions. Within the Bayesian framework, the probabilistic representation of weights via distributions enables the quantification of uncertainty in model predictions. A BNN aims to approximate the distribution of the weights $\mathbf{W}$ as a function $\mathbf{Y}=f_{\mathbf{W}}(\mathbf{X})$ that represents the neural network model. A prior distribution is assigned to the weights $p(\mathbf{W})$ that represents the initial belief of the weights before observing any data. The goal is to infer the posterior distribution of the weights $p(\mathbf{W}|\mathbf{X}, \mathbf{Y})$ using Bayes law:
\begin{equation}
\label{eq:2}
p(\mathbf{W}|\mathbf{X},\mathbf{Y}) = \frac{p(\mathbf{Y}|\mathbf{X}, \mathbf{W})p(\mathbf{W})}{p(\mathbf{Y}|\mathbf{X})} = \frac{p(\mathbf{Y}|\mathbf{X}, \mathbf{W}) p(\mathbf{W})}{\int p(\mathbf{Y}|\mathbf{X}, \mathbf{W}) p(\mathbf{W}) d\mathbf{W}},
\end{equation}
where $p(\mathbf{Y}|\mathbf{X})$ is the evidence factor and $p(\mathbf{Y}|\mathbf{X}, \mathbf{W})$ is the model likelihood. 

The integral in Eq.(\ref{eq:1}) is estimated by sampling $T$ distinct weight sets $w_i$ from the posterior distribution using the Monte Carlo method, corresponding to an ensemble of $T$ predictions generated by the BNN. For this study, we chose $T = 100$ to balance accuracy and complexity, following \cite{Filos2019} and \cite{Ortiz2023}. The final prediction $\mathbf{\hat{Y}}$, given the inputs $\mathbf{X}^{*}$, is calculated as the ensemble mean of T (here 100) predictions:
\begin{equation}
\label{eq:3}
\mathbf{\hat{Y}} =  \frac{1}{T} \sum_{i=1}^{T} p(\mathbf{Y}^*|\mathbf{X}^*, w_i)  \approx p(\mathbf{Y}^*|\mathbf{X}^*), \quad w_i \sim p(\mathbf{W} | \mathbf{X}, \mathbf{Y}).
\end{equation}
However, the posterior distribution $p(\mathbf{W}|\mathbf{X}, \mathbf{Y})$ is often intractable because the evidence factor $p(\mathbf{Y}|\mathbf{X})$ cannot be analytically evaluated for high-dimensional neural networks \citep{Blei2017}. To address this issue, our study employs two popular approaches that enable analytically tractable inference: variational inference \citep{Graves2011, Ranganath2014, Blundell2015} and Monte Carlo dropout \citep{Gal2016}.

Variational inference approximates the posterior distribution $p(\mathbf{W}|\mathbf{X}, \mathbf{Y})$ with a simple distribution $q_{\theta}(\mathbf{W})$, parameterized by variational parameters $\theta$. The goal is now turned from an inference problem into an optimization problem, identifying the variational distribution $q_{\theta}(\mathbf{W})$ that is closest to the target posterior distribution $p(\mathbf{W}|\mathbf{X}, \mathbf{Y})$. The Kullback‐Leibler (KL) divergence between $q_{\theta}(\mathbf{W})$ and $p(\mathbf{W}|\mathbf{X}, \mathbf{Y})$, $KL[q_{\theta}(\mathbf{W})||p(\mathbf{W}|\mathbf{X}, \mathbf{Y})]$, measures how $q_{\theta}(\mathbf{W})$ diverges from $p(\mathbf{W}|\mathbf{X}, \mathbf{Y})$. Thus, the task becomes the problem of minimizing the KL divergence, which is equivalent to maximizing the Evidence Lower BOund \citep[ELBO;][]{bishop2006pattern} loss:
\begin{equation}
\label{eq:4}
\mathcal{L} \equiv \int q_{\theta}(\mathbf{W}) \log p(\mathbf{Y}|\mathbf{X}, \mathbf{W}) d\mathbf{W} - \pi KL[q_{\theta}(\mathbf{W})||p(\mathbf{W})],
\end{equation}
where the first term measures the model's fit to the observed data and the second term is a measure of how close the variational distribution $q_{\theta}(\mathbf{W})$ is to the prior distribution $p(\mathbf{W})$. The coefficient $\pi$ determines the effect of the prior on the ELBO loss. 

In mean field variational inference, the variational distribution of weights $q_{\theta}(\mathbf{W})$ is modeled with a Gaussian distribution $\mathcal{N}(\mu, \sigma^2)$ with mean $\mu$ and variance $\sigma^2$, and each weight is independently sampled from the distribution. The parameters $\mu$ and $\sigma$ are learned through optimizing the ELBO loss during training. The prior distribution $p(\mathbf{W})$ is usually assumed to be an independent Gaussian distribution $\mathcal{N}(\mu_0, {\sigma_0}^2)$. In our study, we propose two approaches to set the prior distribution $p(\mathbf{W})$ for BNNs. The first approach (Fig. \ref{fig2}c) is to adopt the commonly used Gaussian distribution $\mathcal{N}(0, 1)$. This approach (Fig. \ref{fig3}b) searches for a posterior distribution of hypotheses $h_{Bayesian},$ rather than a single hypothesis for the deterministic ResNet, within a randomly chosen prior distribution of hypotheses $h_{prior}$ in the hypothesis space $\mathcal{H}$. Compared to a point hypothesis, a posterior distribution of hypotheses is likely to capture the underlying characteristics within the data more effectively and perform better on unseen data. A sensitivity experiment on $\pi$ determined that $\pi=1/3$ optimizes the performance of the BNN on the validation dataset at the 10-minute lead time and is applied to all lead times. In the second approach (Fig. \ref{fig2}b), we build the BNN with the MOdel Priors with Empirical Bayes using Deep neural network \citep[MOPED;][]{Krishnan2020} method to initialize the prior distribution using the pretrained weights, $w_d$, from the deterministic ResNet to learn the posterior distribution, integrating strengths from both classical (frequentist) and Bayesian statistical frameworks. This approach (Fig. \ref{fig3}c) searches for a posterior distribution hypotheses $h_{Bayesian-MOPED}$ in an informed prior distribution of hypotheses $h_{Det-prior}$ near the deterministic hypothesis $h_{d}$. Specifically, the mean $\mu_0$ is taken from the pretrained weights $w_d$ and the scale $\sigma_0$ is set to $0.1|w_d|$. $\pi=1$ is used to allocate equal weight to the influence of the prior and the effect of data fitting.

In our study, we adapted deterministic ResNet to Bayesian ResNet with the above variational inference methods using Tensorflow and Tensorflow-Probability \citep{Dillon2017}. The changes from deterministic to Bayesian ResNet are highlighted in orange and purple in Fig. \ref{fig2}b-c. Specifically, we replaced the deterministic ResNet convolutional and dense layers with variational layers for the Bayesian ResNet (Fig. \ref{fig2}c) and its variant with MOPED (Fig. \ref{fig2}b). The Reparameterization \citep{Kingma2015} and Flipout \citep{Wen2018} methods are commonly used to estimate the parameters of the posterior distribution in variational layers by sampling the distribution $q_{\theta}(\mathbf{W})$. The Reparameterization method applies a uniform gradient of loss with respect to parameters across all training samples in a minibatch, leading to correlated gradients and a higher variance in gradient estimates. In contrast, the Flipout method introduces random perturbations (flips) to the gradient for each sample within the mini-batch, effectively decorrelating the gradients in a minibatch and reducing variance. Because of its lower variance, We chose the Flipout method rather than the Reparameterization method. In Bayesian ResNet with MOPED (Fig. \ref{fig2}b), the weights of variational and batch normalization layers are initialized from the established weights of the corresponding layers in the deterministic ResNet. For both Bayesian ResNet models, ensemble predictions are produced by running the model 100 times, with weights randomly sampled from the variational distribution each time.

A viable alternative to approximate the posterior distribution is Monte Carlo (MC) dropout (Fig. \ref{fig2}d). \cite{Gal2016} established that MC dropout, i.e., random deactivation of neurons, in the inference stage is equivalent to variational inference. In contrast to the variational inference approach, models using MC dropout do not include an explicit representation of the posterior weight distribution. However, applying dropout in the inference stage is an approximation of sampling the posterior distribution and can generate a unique prediction. The deterministic ResNet + MC dropout (Fig. \ref{fig3}d) produces an ensemble of hypotheses $h_{dMC}$ in the neighborhood of $h_{d}$ by randomly deactivating neurons in the inference stage. For our study, we generated the ensemble predictions by running the deterministic ResNet 100 times with MC dropout activated. The dropout rate of the deterministic ResNet in \cite{Fan2024} is used.

\subsubsection{INITIAL-WEIGHTS ENSEMBLES}
The initial-weights ensemble, often referred to as the "deep ensemble" \citep{Lakshminarayanan2017}, is usually considered non-Bayesian, but is considered as a BMA approach when providing an accurate estimate of the predictive distribution through functional diversity \citep{Wilson2020}. Rather than modeling the weight distribution as in BNNs, the initial-weights method (Fig. \ref{fig2}e) trains an ensemble of models with the same architecture and data splits but different random initial weight sets. This method (Fig. \ref{fig3}e) generates an ensemble of hypotheses, $h_{DEn}$, from deterministic models with different initial weight sets, each setting a unique direction of approach within the hypothesis space $\mathcal{H}$ as they converge toward the ground truth $f^{*}$. In our study, initial-weights ensembles were constructed by training 100 deterministic ResNet models.

The initial-weights ensemble + MC dropout (Fig. \ref{fig2}f) applies MC dropout to initial-weights ensembles. This approach (Fig. \ref{fig3}f) produces an ensemble of hypotheses $h_{DEnMC}$ in proximity to $h_{DEn}$. Results of the initial-weights ensemble + MC dropout were generated by collecting the forecasts of 100 deterministic ResNet models with MC dropout activated in the inference stage.



\subsection{Model evaluation}
To evaluate the forecast skill of these methods, we use the Brier Score \citep[BS;][]{Brier1950-af} and Brier Skill Score \citep[BSS;][]{Wilks2019} over the testing dataset. The BS is a measure of forecast errors using the mean squared difference between the observed label, $y_{j,k}$, and the forecast probability, $p_{j,k}$, given by
\begin{equation}
\label{eq:5}
BS = \frac{1}{NK} \sum_{j=1}^{N} \sum_{k=1}^{K} (y_{j,k} - p_{j,k})^2,
\end{equation}
where $N$ is the total number of events and K is the total number of classes. The BSS is calculated as $BSS = 1 - \frac{BS_{\text{forecast}}}{BS_{\text{climatology}}}$, where $BS_{\text{forecast}}$ is the BS of the predicted forecasts and $BS_{\text{climatology}}$ is the BS of the climatological forecasts. The BSS measures the improvement of forecast skill relative to climatology. The BSS components—reliability and resolution—are also calculated to characterize their differences. The resolution measures the difference between conditional event frequencies and the observed climatological frequency, whereas the reliability measures how close the forecast probabilities are to the observed event frequency at the corresponding probabilities. The reliability score also indicates how well forecast probabilities are calibrated through uncertainty estimates.

Uncertainty is measured by the variance of the ensemble forecasts and is evaluated using spread-skill diagrams and discard tests. The spread-skill diagram compares the prediction error, measured by the BS, against the uncertainty, measured by the variance of the ensemble forecasts. The discard tests \citep{Haynes2023} are employed to evaluate the ranking quality of uncertainty, measured by the standard deviation. They show how prediction errors, measured by the BS, change with the fraction of highest-uncertainty cases discarded. The average discard improvement (DI) score is derived from the discard tests; it is a measure of average performance improvement when the discard fraction is increased.

\section{Results}

\subsection{Brier Skill Score}
\begin{figure*}[h]
\centerline{\includegraphics[width=0.8\textwidth,angle=0]{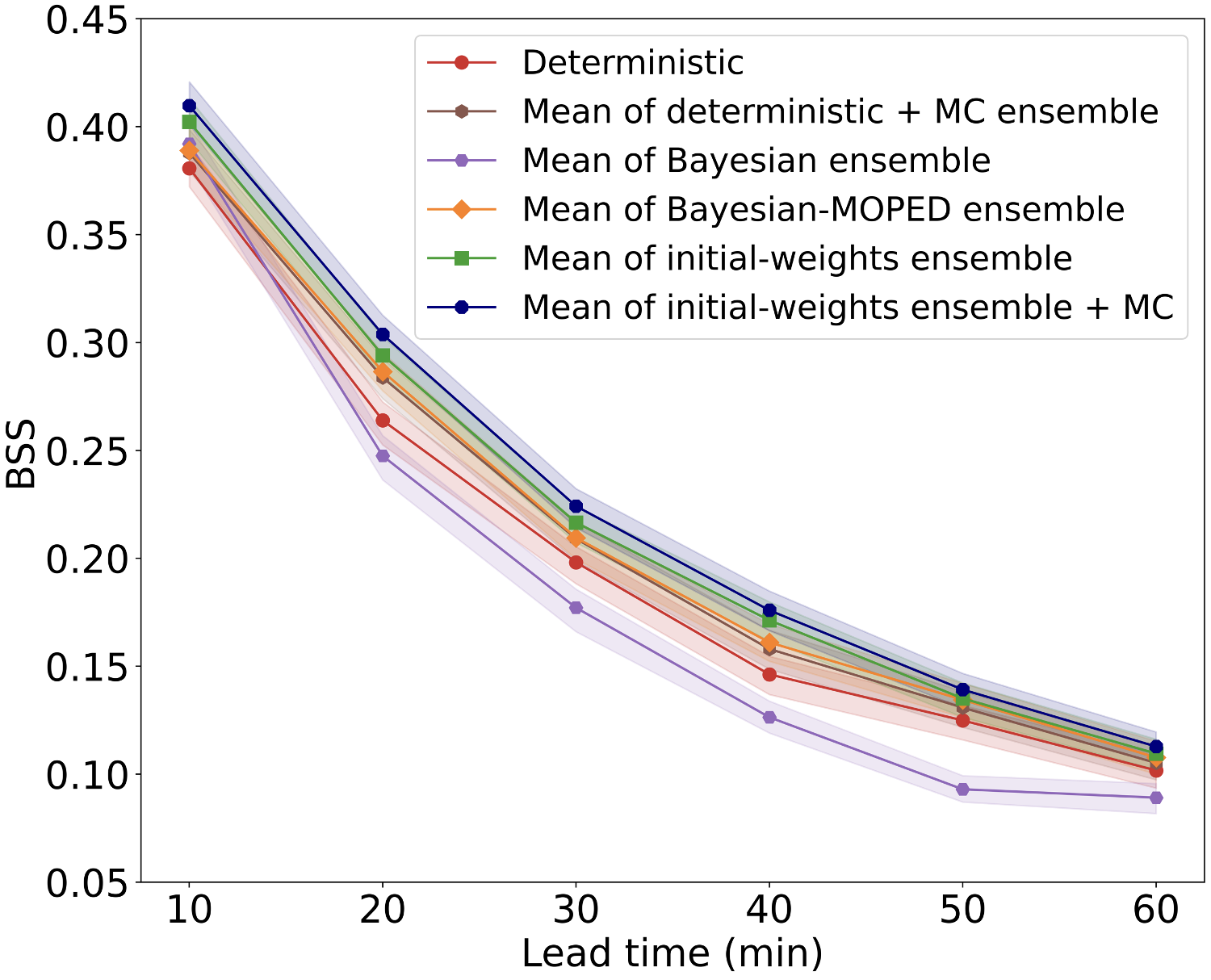}}
\caption{BSSs for the deterministic ResNet, mean of deterministic ResNet + MC dropout ensemble, mean of Bayesian ResNet ensemble, mean of Bayesian-MOPED ResNet ensemble, mean of initial-weights ensemble of ResNets, and mean of initial-weights ensemble of ResNets + MC dropout at lead times from 60 min to 10 min in 10-min steps. Light shading around each curve shows the 95\% confidence intervals determined by bootstrapping the testing samples 1000 times.}\label{fig4}
\end{figure*}

The BSSs for all methods investigated decrease with increasing lead times, as expected, and the ensemble means of all Bayesian methods outperform the deterministic ResNet at the 10-min lead time (Fig. \ref{fig4}). Specifically, the initial-weights ensemble + MC dropout performs the best, followed by the initial-weights ensemble. The deterministic ResNet + MC dropout ensemble, Bayesian ensemble, and Bayesian-MOPED ensemble generate comparable BSSs, outperforming the deterministic ResNet yet performing slightly worse than the two initial-weights ensembles. At lead times longer than 10 min, the initial-weights ensemble + MC dropout still performs the best, whereas the performance of the Bayesian-MOPED ensemble remains close to the deterministic ResNet + MC dropout ensemble, which is slightly better than the deterministic ResNet. However, the Bayesian ensemble performs worse than the deterministic ResNet, likely due to the convergence issue related to BNNs \citep{Schreck2023, Haynes2023}. Note that the hyperparameters of the Bayesian ResNet were optimized specifically for the 10-min lead time forecasts and subsequently directly applied to longer lead times. This reveals that Bayesian models might require additional hyperparameter tuning to prevent convergence to inferior solutions compared to deterministic models. Across all lead times, the BSSs of the initial-weights ensemble + MC dropout are 8-20\% higher than the deterministic ResNet.

\begin{figure*}[!ht]
\centerline{\includegraphics[width=\textwidth,angle=0]{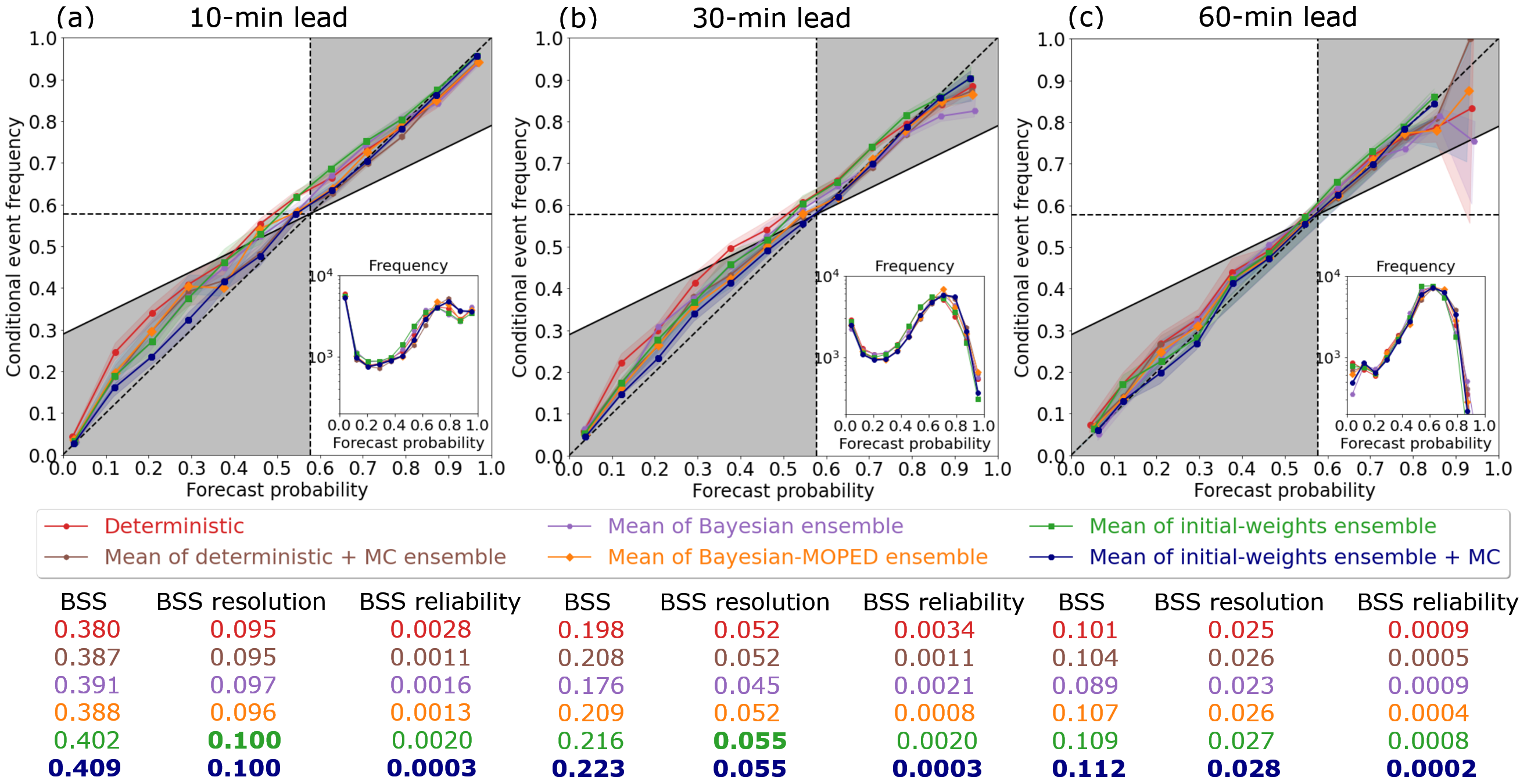}}
\caption{(top) Attribute diagrams and (bottom) BSSs for the deterministic ResNet, mean of deterministic ResNet + MC dropout ensemble, mean of Bayesian ResNet ensemble, mean of Bayesian-MOPED ResNet ensemble, mean of initial-weights ensemble of ResNets, and mean of initial-weights ensemble of ResNets + MC dropout on the testing dataset at (a) 10-, (b) 30-, and (c) 60-min lead times. Light shading around each curve shows the 95\% confidence intervals determined by bootstrapping the testing samples 1000 times. The diagonal dashed line indicates perfect reliability, and the horizontal dashed line represents the climatological event frequency. The gray shaded areas indicate regions where points on the curves produce positive BSSs, whereas the white areas indicate regions where points on the curve generate negative BSSs. The inset panel shows the binned frequencies of the forecast probabilities for each model.}\label{fig5}
\end{figure*}

At 10-, 30-, and 60-min lead times, the reliability curves of the initial-weights ensemble + MC dropout are the closest to the perfect reliability diagonal in Fig. \ref{fig5} among all methods, which explains its lowest reliability scores overall. The initial-weights ensemble + MC dropout also generates more high-probability forecasts than the other methods, except for the Bayesian ensemble, which contributes to its higher resolution scores. The higher resolution scores of the initial-weights ensemble + MC dropout are attributable to the initial-weights ensemble method itself given the comparable resolution scores generated by the two initial-weights ensembles at all lead times. 


Both the deterministic ResNet + MC dropout ensemble and initial-weights ensemble + MC dropout show significantly lower reliability scores than the deterministic ResNet and the initial-weights ensemble, respectively, indicating that the MC dropout method effectively calibrates the probability distributions (Fig. \ref{fig5}). At the 10-min lead time, the Bayesian ensemble shows improved resolution and reliability scores, comparable to those of the deterministic ResNet + MC dropout ensemble and the Bayesian-MOPED ensemble. However, at 30- and 60-min lead times, the resolution scores of the Bayesian ensemble are much lower than those for the deterministic ResNet, deterministic ResNet + MC dropout ensemble, and Bayesian-MOPED ensemble. This suggests that the convergence issue associated with Bayesian models mainly affects their ability to separate conditional and climatological event frequency. The Bayesian-MOPED and deterministic ResNet + MC dropout ensembles show comparable abilities in reducing the reliability score. For the Bayesian-MOPED ensemble, forecast skill may be limited by the constrained search area within the hypothesis space, which is restricted by the prior weights from the deterministic ResNet. Given the similar performances of the Bayesian-MOPED ensemble and deterministic ResNet + MC dropout ensemble, together with the poor performance of the Bayesian ensemble at longer lead times, subsequent analyses herein will concentrate on the performance of the Bayesian-MOPED ensemble, initial-weights ensemble, and initial-weights ensemble + MC dropout relative to the deterministic ResNet and to each other.

\clearpage
\begin{figure*}[!ht]
\centerline{\includegraphics[width=0.79\textwidth,angle=0]{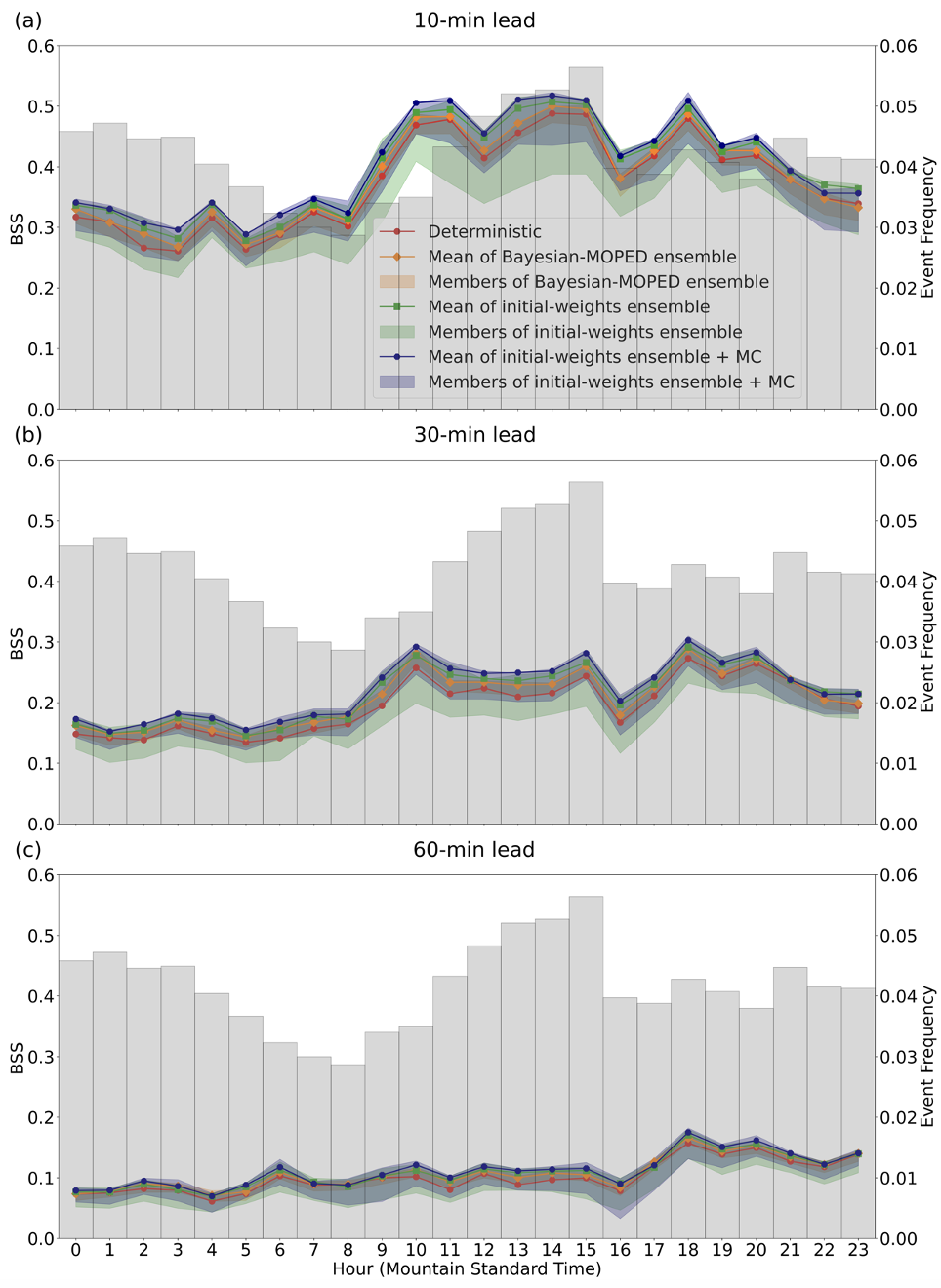}}
\caption{Hourly BSS comparisons for the deterministic ResNet, Bayesian-MOPED ResNet ensemble, initial-weights ensemble of ResNets, and initial-weights ensemble of ResNets + MC on the testing dataset at (a) 10-, (b) 30-, and (c) 60-min lead times. Solid lines represent the BSS of the deterministic and the ensemble mean of Bayesian methods. Shaded areas represent the BSSs of individual members. The grey histogram represents event frequency.}\label{fig6}
\end{figure*}

BSSs for the deterministic ResNet, Bayesian-MOPED ensemble, initial-weights ensemble, and initial-weights ensemble + MC dropout at 10-, 30-, and 60-min lead times are shown in Fig. \ref{fig6} for different hours during the diurnal cycle. The mean BSS is calculated using the mean probabilistic forecast over all members. At all lead times, the mean of the initial-weights ensemble + MC dropout outperforms the other methods at nearly all hours. The ensemble means of all Bayesian methods show BSSs superior to that of most individual members, indicating that the ensemble mean method substantially improves forecast skill over deterministic predictions, consistent with \cite{Sha2024}. 

At the 10- (Fig. \ref{fig6}a) and 30-min (Fig. \ref{fig6}b) lead times, all methods show high BSSs from late morning (10:00 MST) through early evening (20:00 MST) and low BSSs from midnight (00:00 MST) into early morning (08:00 MST). Performance of the mean of the initial-weights ensemble + MC dropout surpasses that of the initial-weights ensemble primarily between 10:00 MST and 15:00 MST. At the 60-min lead time, all methods exhibit higher BSSs from early evening (18:00 MST) until midnight (00:00 MST) than for other hours. The mean of the initial-weights ensemble + MC dropout produces slightly higher BSSs than the mean of the initial-weights ensemble and Bayesian-MOPED ensemble. Poor forecast skill around midnight is likely due to the inherent challenges in predicting nocturnal convection \citep[e.g.,][]{Johnson2017, Blake2017}. 

At all lead times (Fig. \ref{fig6}), both the initial-weights ensemble and initial-weights ensemble + MC dropout show a larger ensemble spread than the Bayesian-MOPED ensemble, suggesting that the improved forecast skill by the initial-weights ensembles may be attributable to the representation of multiple solutions and a more extensive search in the hypothesis space (Fig. \ref{fig3}e and \ref{fig3}f).

\begin{figure*}[h]
\centerline{\includegraphics[width=\textwidth,angle=0]{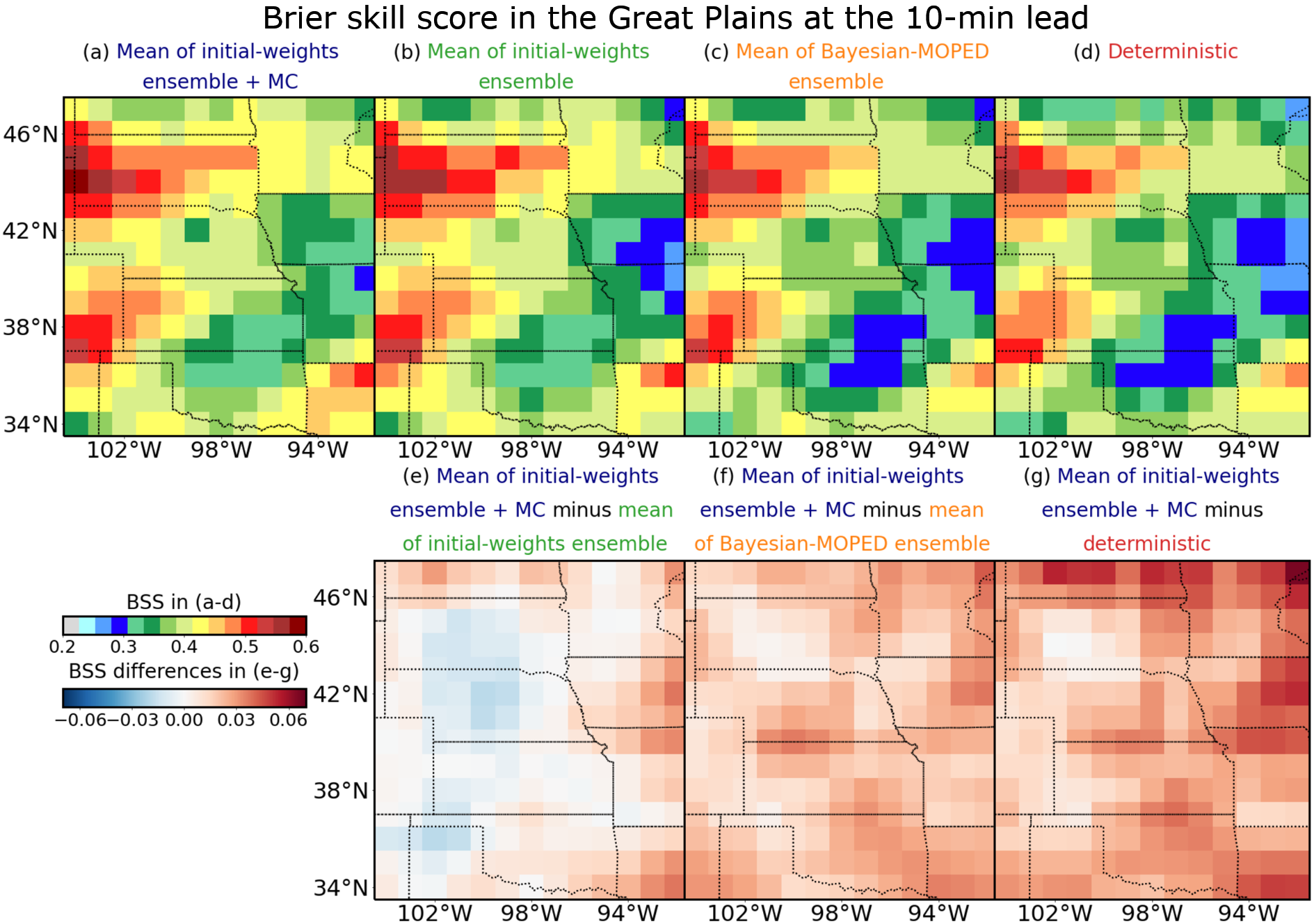}}
\caption{Grid-point-wise BSSs at a 1$^{\circ}$ grid spacing for the means of the (a) initial-weights ensemble of ResNets + MC dropout, (b) initial-weights ensemble of ResNets, (c) mean of Bayesian-MOPED ResNet ensemble, and (d) deterministic ResNet at the 10-min lead time. (e-g) Differences between (a) and (b), (a) and (c), and (a) and (d), respectively. For each grid point, events over the 3$\times$3 grid cells centered on the grid point are included in calculation of the grid-point-wise BSSs.}\label{fig7}
\end{figure*}

\begin{figure*}[h]
\centerline{\includegraphics[width=\textwidth,angle=0]{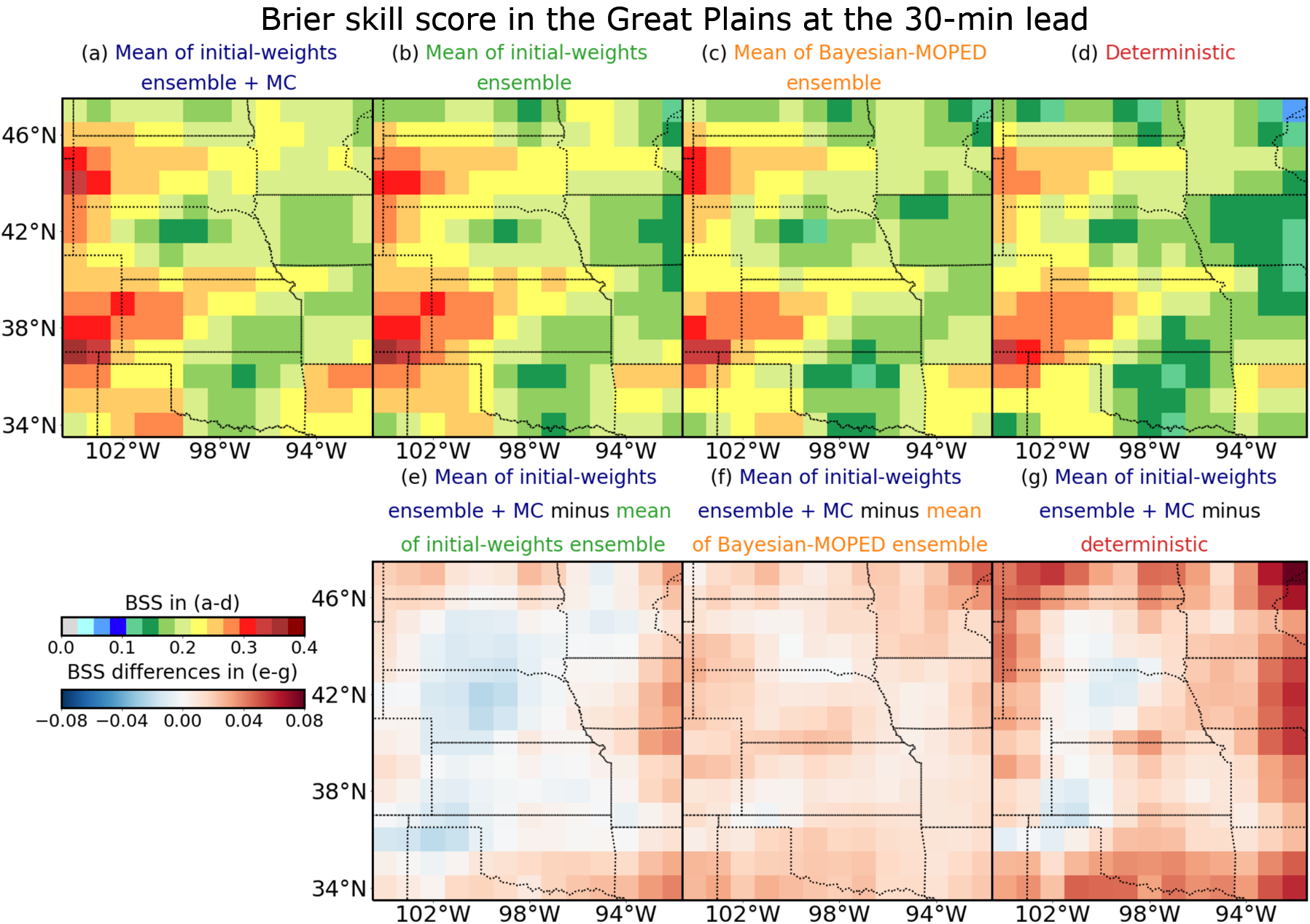}}
\caption{As in Fig. \ref{fig7}, but for the 30-min lead time.}\label{fig8}
\end{figure*}

\begin{figure*}[h]
\centerline{\includegraphics[width=\textwidth,angle=0]{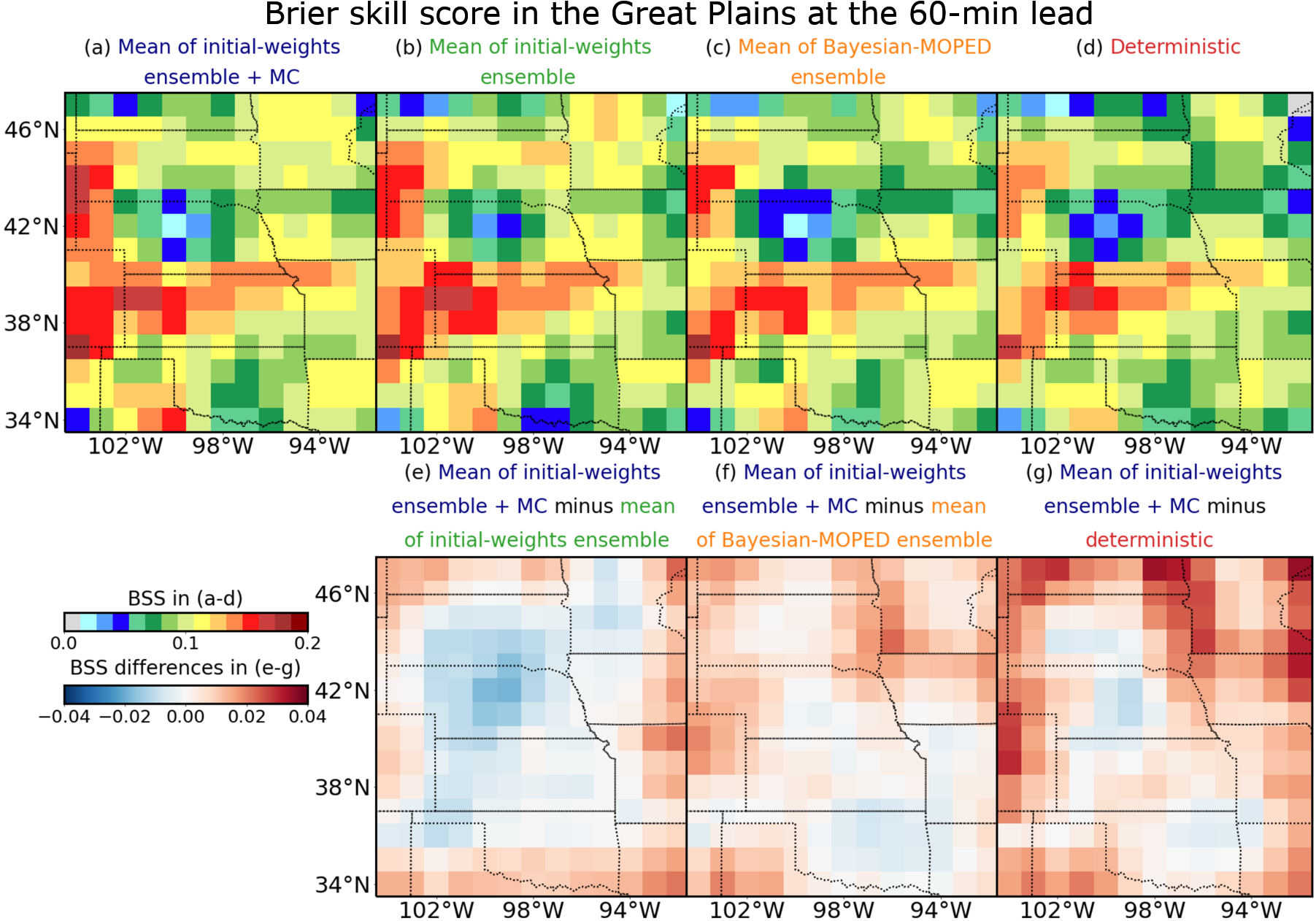}}
\caption{As in Fig. \ref{fig7}, but for the 60-min lead time.}\label{fig9}
\end{figure*}

Spatial distributions of grid-point-wise BSSs at a 1$^{\circ}$ grid spacing are shown in Figs. \ref{fig7}-\ref{fig9} for the deterministic ResNet, Bayesian-MOPED ensemble, initial-weights ensemble, and initial-weights ensemble + MC dropout at 10-, 30-, and 60-min lead times. For each grid point, testing data for all 3$\times$3 grid points centered on it are included to increase the sample size for the BSS calculations.

At all lead times, all methods perform well in eastern Colorado, western Kansas, and eastern South Dakota. In these regions, CI is triggered by several mechanisms, including orographic lifting, terrain-induced convergence, and synoptic-scale forcing. Within this dynamic environment, all methods successfully capture the cloud-top and moisture characteristics from infrared observations to effectively enhance CI forecast skill. On the other hand, all methods exhibit poor performance in flat terrain regions in eastern Kansas, Iowa, and Missouri, all of which have high event frequencies, along with Oklahoma and Arkansas (Fig. \ref{fig1}b). This suggests that temporally frequent occurrences of CI are challenging to predict with these methods.

At the 10-min lead time, the initial-weights ensemble + MC dropout outperforms the deterministic ResNet in almost all regions, and particularly in North Dakota, Minnesota, Iowa, and Arkansas (Fig. \ref{fig7}g). At 30- and 60-min lead times, the initial-weights ensemble + MC dropout showed lower forecast skill than the deterministic ResNet in some regions, and especially in central Nebraska (Figs. \ref{fig8}g and \ref{fig9}g). The initial-weights ensemble + MC dropout surpasses the Bayesian-MOPED ensemble in almost all regions at 10- and 30-min lead times (Figs. \ref{fig7}f and \ref{fig8}f), while it performs slightly worse in Oklahoma compared to the Bayesian-MOPED ensemble at the 60-min lead time (Fig. \ref{fig9}f). At all lead times, the initial-weights ensemble + MC dropout shows inferior forecast skill to the initial-weights ensemble in South Dakota, Nebraska, and Kansas, but demonstrates better performance in other regions (Fig. \ref{fig7}e, \ref{fig8}e, and \ref{fig9}e).

Note that the testing dataset only contains events in the Great Plains region during June 2021 rather than a longer period and a broader area. Performance might vary significantly across different seasons and regions.

\subsection{Generalization test}
\begin{figure*}[h]
\centerline{\includegraphics[width=\textwidth,angle=0]{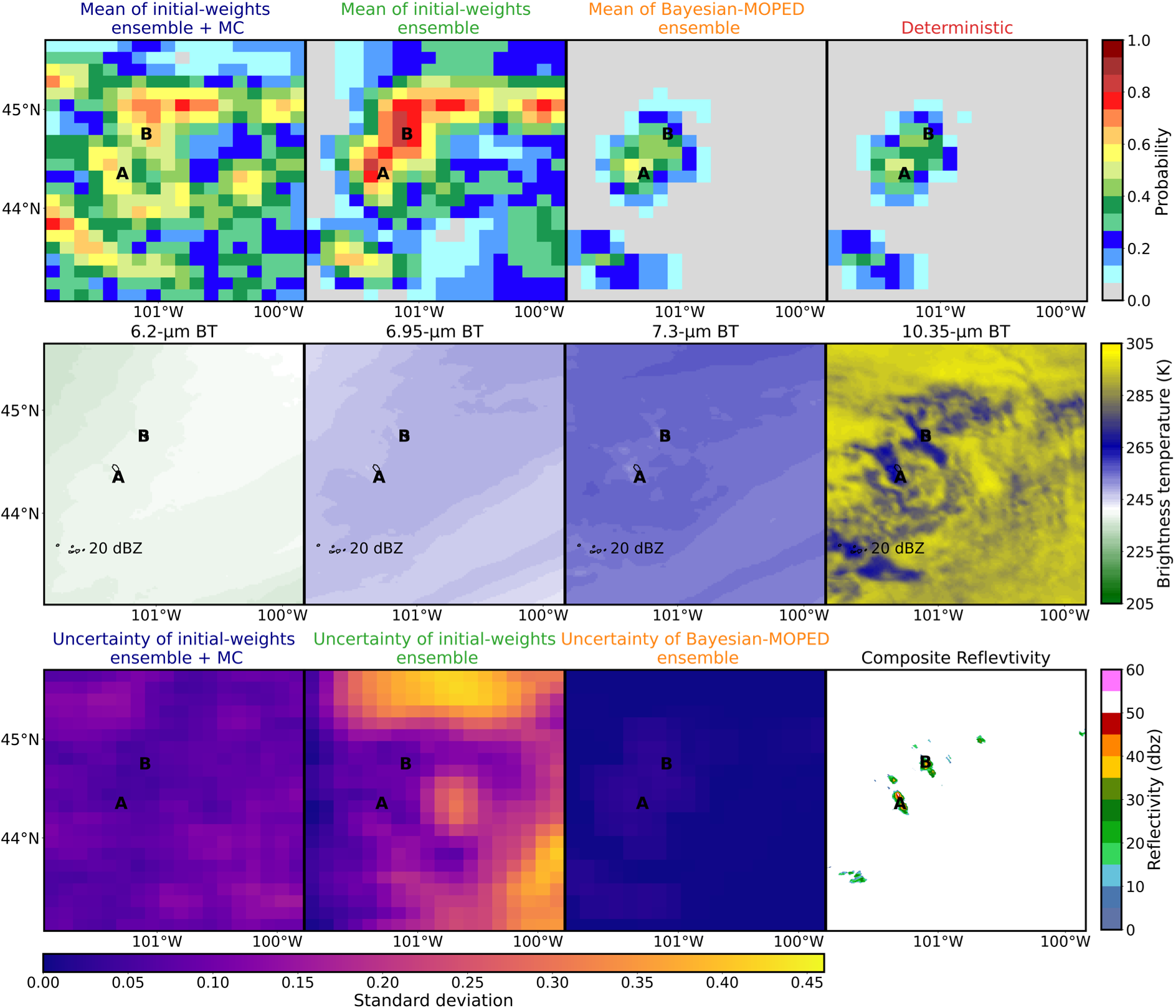}}
\caption{Forecasts for two isolated CI events and their neighboring regions in South Dakota at 16:01 UTC 28 June 2021 at the 10-min lead time. (row 1) Probabilistic forecasts by the means of the initial-weights ensemble of ResNets + MC dropout, initial-weights ensemble of ResNets, Bayesian-MOPED ResNet ensemble, and deterministic ResNet. (row 2) Observed 6.2-, 6.95-, 7.3-, and 10.35-$\mu$m BTs overlaid with black contours of 20 dBZ composite reflectivity at the forecast time. (row 3) The first three images show the uncertainty, as given by the standard deviation, of the initial-weights ensemble of ResNets + MC dropout, initial-weights ensemble of ResNets, and Bayesian-MOPED ResNet ensemble forecasts. The last image shows the composite reflectivity at 16:11 UTC 28 June 2021. The labels "A" and "B" denote the location of the two CI events observed at the same time.}\label{fig10}
\end{figure*}

\begin{figure*}[h]
\centerline{\includegraphics[width=\textwidth,angle=0]{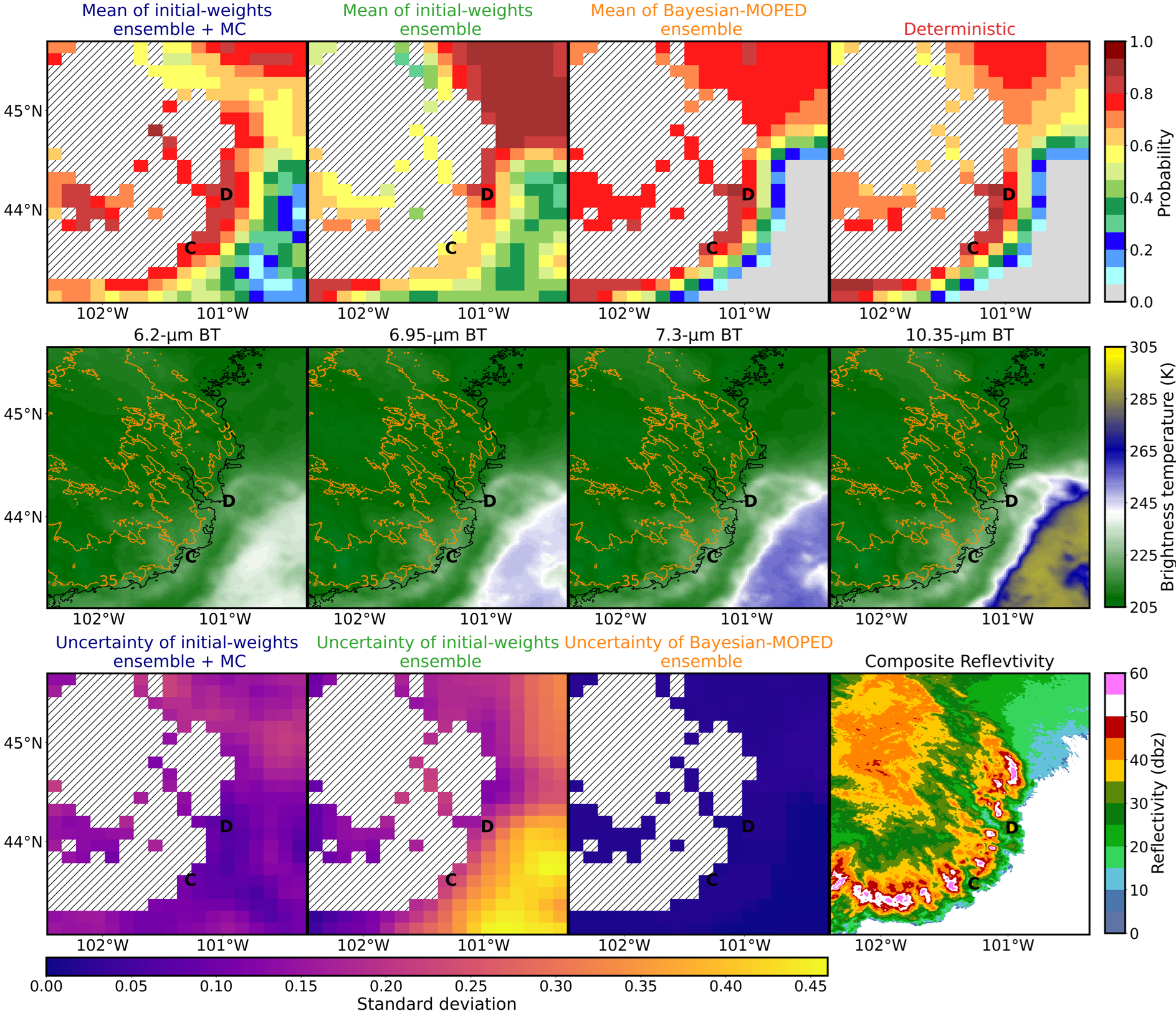}}
\caption{As in Fig. \ref{fig10}, but for the mountainous regions in western South Dakota at 06:23 UTC 9 June 2021 at the 10-min lead time. Grid points with pre-existing convection at the forecast time are marked using hatches. In the second row of images, the black and orange contours represent 20 dBZ and 35 dBZ composite reflectivity, respectively, at the forecast time. The observed CI locations "C" and "D" and composite reflectivity in the last image are at 06:33 UTC 9 June 2021.}\label{fig11}
\end{figure*}

To investigate spatial generalization by these methods, we selected two CI scenarios, isolated CI (Fig. \ref{fig10}) and CI obscured by anvil clouds (Fig. \ref{fig11}). The target region is the 12 km × 12 km central region of each patch (see Fig. \ref{fig1}c).

Figure \ref{fig10} shows CI forecasts, including ensemble mean probability (Fig. \ref{fig10}, row 1) and uncertainty (Fig. \ref{fig10}, row 3), 6.2-, 6.95-, 7.3-, and 10.35-$\mu$m BTs (Fig. \ref{fig10}, row 2), and composite reflectivity (Fig. \ref{fig10}, row 3) in South Dakota. Two isolated CI events, labeled "A" and "B", originated from clear-sky regions. Only weak moisture gradients are observed near the two events in the lower (6.95-$\mu$m BT) and middle (7.3-$\mu$m BT) troposphere. The probability and uncertainty forecasts for each grid point are calculated using the mean forecasts for the point and its eight surrounding grid points. In the region near CI event "A", both the deterministic ResNet and mean of the Bayesian-MOPED ensemble forecasts predict moderate CI probabilities around 0.4, whereas the mean of the initial-weights ensemble + MC dropout forecasts generates a slightly higher probability of about 0.5. The mean probability of the initial-weights ensemble is significantly higher than for the other methods. In the region neighboring event "B", both initial-weights ensembles generate probabilities exceeding 0.6, whereas the deterministic ResNet and mean of the Bayesian-MOPED ensemble have probabilities below 0.3. Overall, the two initial-weights ensembles show better forecast skill than the other methods, suggesting that they are more effectively capturing the signals from the moisture distribution in the troposphere. However, the two initial-weights ensembles exhibit slightly higher uncertainty for events "A" and "B" compared to the Bayesian-MOPED ensemble. 

For the rest of the clear-sky regions without occurrence of CI, the two initial-weights ensembles demonstrate poor generalization, particularly in the regions characterized by low 10.35-$\mu$m BTs. In contrast, the deterministic ResNet and the mean of the Bayesian-MOPED ensemble forecasts exhibit good generalization, generating low probabilities in most regions. The initial-weights ensemble shows a significantly higher uncertainty than the other ensemble methods in some areas.

Figure \ref{fig11} presents CI forecasts, BTs, and composite reflectivity for the mountainous regions in western South Dakota. Two CI events, labeled "C" and "D", developed under optically thick anvil clouds. Those regions with preexisting convection at the forecast time are marked with hatches and have been excluded from the analysis based on the CI criteria above and in \cite{Fan2024}. For both events, nearly all methods predict high CI probabilities above 0.8, except for the 0.6 probability for event "C" by the mean of the initial-weights ensemble forecasts. 

In the clear-sky region at the lower right of the scene, the deterministic ResNet and Bayesian-MOPED ensemble show better generalization compared to the two initial-weights ensembles. In the anvil regions at the upper right of the scene, all methods exhibit poor generalization, with the deterministic ResNet and initial-weights ensemble + MC dropout showing slightly better skill than the initial-weights ensemble and Bayesian-MOPED ensemble. In all regions, the initial-weights ensemble shows higher uncertainties than the other ensembles.

Overall, compared to the deterministic ResNet and Bayesian-MOPED ensemble, the two initial-weights ensembles exhibit better forecast skill for CI events in clear-sky regions but show weaker spatial generalization over other non-CI clear-sky regions. In regions covered by anvil clouds, though all methods demonstrate good forecast skill for CI events, they show poor spatial generalization over the rest of the non-CI regions.

\subsection{Spread-skill diagram and discard tests}
\begin{figure*}[h]
\centerline{\includegraphics[width=\textwidth,angle=0]{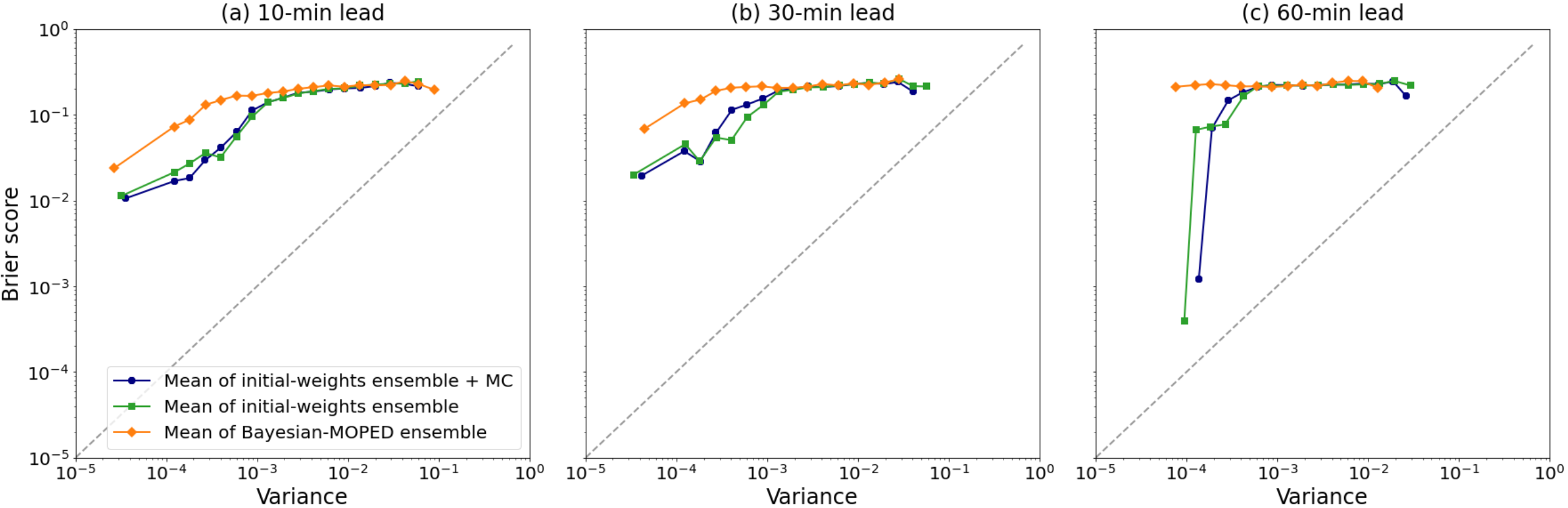}}
\caption{Spread-skill diagrams for forecasts from the means of the Bayesian-MOPED ensemble, initial-weights ensemble, and initial-weights ensemble + MC dropout methods at (a) 10-, (b) 30-, and (c) 60-min lead times. The dashed diagonal lines indicate the 1-to-1 lines.}\label{fig12}
\end{figure*}
We employed spread-skill diagrams and discard tests to analyze how the forecast skill of these Bayesian methods depends on ensemble spread and uncertainty. The spread-skill diagrams (Fig. \ref{fig12}) illustrate the relationship between the ensemble spread and the forecast errors. At all lead times, the two initial-weights ensembles show comparable BSs that are significantly lower than for the Bayesian-MOPED ensemble when the variance is lower than $10^{-3}$. However, the BSs of all methods converge when the variance exceeds $10^{-3}$. This suggests that the two initial-weights ensembles are more skillful at low uncertainties compared to the Bayesian-MOPED ensemble, whereas all methods perform comparably at high uncertainties.

\begin{figure*}[h]
\centerline{\includegraphics[width=\textwidth,angle=0]{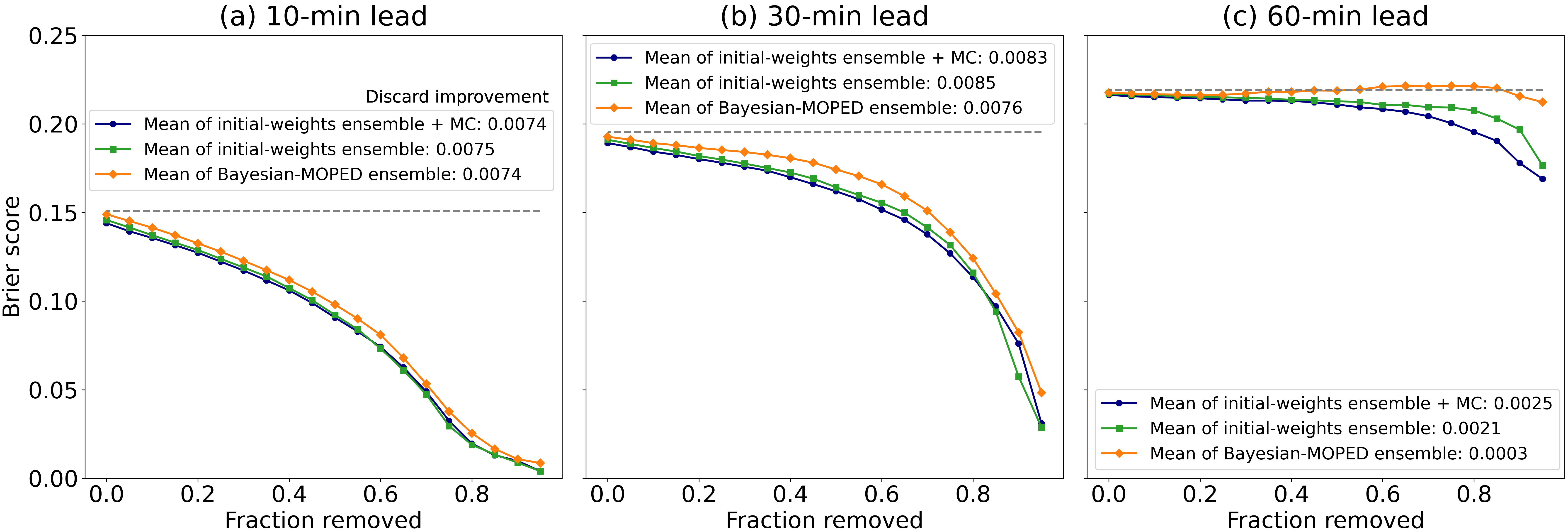}}
\caption{Discard test curves for the means of the initial-weights ensemble of ResNets + MC dropout, initial-weights ensemble of ResNets, and Bayesian-MOPED ResNet ensemble at (a) 10-, (b) 30-, and (c) 60-min lead times. The horizontal gray dashed lines indicate the BSs of the deterministic ResNet for reference. The legend in each panel shows the discard improvement (DI; higher is better) score.}\label{fig13}
\end{figure*}
The discard tests (Fig. \ref{fig13}) evaluate the ranking quality of uncertainty by examining how prediction errors, measured by the BS, change with the fraction of highest-uncertainty cases discarded. The overall negative slopes across most fraction bins indicate uncertainty is well-calibrated and effectively separates forecasts with large and small errors across all lead times for all Bayesian methods. A steeper error reduction rate indicates a better ranking quality of uncertainty. According to the discard improvement score, a measure of the average reduction rate, all Bayesian methods perform comparably at the 10-min lead time, with both initial-weights ensembles outperforming the Bayesian-MOPED ensemble at the 30- and 60-min lead times. 


\section{Discussion}
BNNs learn full distributions of model parameters and capture uncertainty from both data variability and model architecture. However, BNNs require optimizing more weights than for deterministic methods and BNNs usually encounter challenges in converging to an optimal solution with skillful performance. The attribute diagrams in Fig. \ref{fig5} indicate that the Bayesian ensemble, when fine-tuned with optimal hyperparameters at the 10-min lead time, outperforms the deterministic ResNet at the 10-min lead time, but its performance is inferior to the deterministic ResNet at longer lead times, likely due to convergence issues.

\begin{figure*}[h]
\centerline{\includegraphics[width=\textwidth,angle=0]{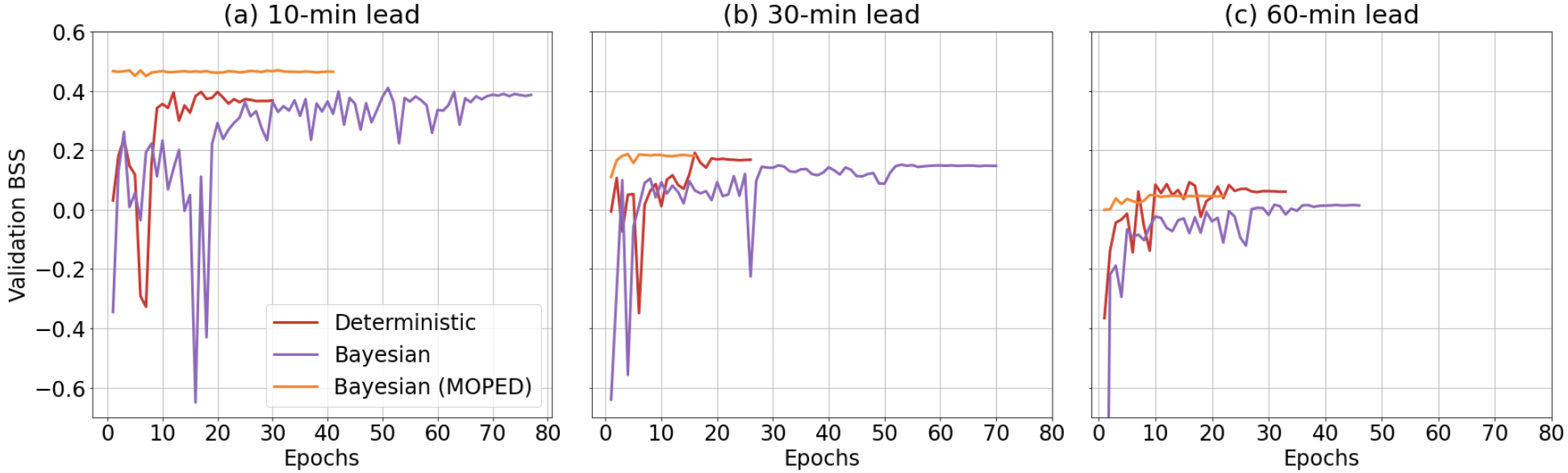}}
\caption{Training convergence curves for the deterministic ResNet, Bayesian ResNet ensemble, and Bayesian-MOPED ResNet ensemble at (a) 10-, (b) 30-, and 60-min lead times.}\label{fig14}
\end{figure*}

Figure \ref{fig14} shows the training convergence curves for the deterministic ResNet, Bayesian ensemble ResNets, and Bayesian-MOPED ensemble ResNets at 10-, 30-, and 60-min lead times. At all lead times, the Bayesian ensemble ResNet spent many more epochs to converge compared to the deterministic and Bayesian-MOPED ensemble ResNets, with the Bayesian-MOPED ensemble ResNets achieving convergence within 10 epochs. Note that the Bayesian-MOPED ensemble ResNets exhibit considerably fewer fluctuations during training than the Bayesian ensemble ResNets, likely resulting from the constraint imposed by the prior weights from the deterministic ResNet. Thus, the MOPED method largely accelerates and stabilizes training convergence for the BNN. The higher BSSs of the Bayesian-MOPED ensemble compared to the Bayesian ensemble at longer lead times (Fig. \ref{fig4}) indicates that the MOPED method also improves forecast skill of the BNN. These findings are consistent with the results in previous studies \citep{Krishnan2020,Zhang2022,Milanes-Hermosilla2023}. Note that the validation BSS of the deterministic ResNet is higher than that of the Bayesian-MOPED ensemble, while the Bayesian-MOPED ensemble generates a higher BSS on the testing data compared to the deterministic ResNet.

\cite{Krishnan2020} evaluated Bayesian and Bayesian-MOPED models based on different model architectures for classification problems. For complex architectures (e.g., ResNet-101, C3D, and VGG), the Bayesian model initialized from random priors struggled to converge to a solution with a performance comparable to the deterministic model, whereas the Bayesian-MOPED models effectively addressed the issue and achieved comparable or better performance compared to the deterministic model. \cite{Zhang2022} demonstrated that the informed priors of the MOPED methods provide a reasonable search area for an optimal solution, thereby guiding the training process towards the minima of basins with less computational costs and better model generalization. \cite{Milanes-Hermosilla2023} also suggested that MOPED methods improve the forecast skill of Bayesian methods to be comparable to, or better than, the deterministic method when applied to motor imagery classification.

\section{Conclusions}
We evaluated several recently proposed Bayesian deep learning methods, including Bayesian and initial-weights ensemble methods, on CI forecast skill up to 1 hour lead times using GOES-16 satellite infrared observations. The BSSs of the Bayesian methods were compared against the deterministic ResNet. Most Bayesian methods outperformed the deterministic ResNet across all lead times, with the initial-weights ensemble + MC dropout performing the best in overall, hourly, and spatial BSSs, as well as probability calibration. The Bayesian ensemble was the only method to perform worse than the deterministic ResNet at the two longer lead times, likely due to the challenge of optimizing a larger number of parameters. To address this issue, we tested the Bayesian-MOPED method, which enhanced the forecast skill of the Bayesian ensemble by constraining the hypothesis search near the deterministic hypothesis. Training convergence curves suggest that the MOPED method accelerated and stabilized training convergence and improved the forecast skill of the BNN. Both initial-weights ensemble methods likely benefited from generating multiple solutions that more thoroughly sampled the hypothesis space.

We also assessed the uncertainty quality and calibration of the three best-performing Bayesian methods through spread-skill diagrams and discard tests. The Bayesian-MOPED ensemble and two initial-weights ensembles demonstrated well-calibrated uncertainty and effectively distinguished between forecasts with large and small errors across all lead times. Among them, both initial-weights ensembles performed slightly better than the Bayesian-MOPED ensemble.

We further evaluated generalization of the Bayesian methods on two selected CI scenarios, isolated CI in clear skies and CI obscured by anvil clouds, and their neighboring regions. Compared to the other methods, the two initial-weights ensembles showed better forecast skill for the selected CI events in clear-sky regions, but exhibited weaker spatial generalization over the non-CI clear-sky regions. In anvil cloud regions, all Bayesian methods demonstrated skillful forecasts for the selected CI events but showed poor generalization over the rest of the non-CI regions.

\clearpage
\acknowledgments
This research benefited from the National Center for Atmospheric Research’s (NCAR’s) Computational and Information Systems Laboratory visitor program. The machine learning training, evaluation, and explanation were performed on an Institute for Computational and Data Sciences supercomputer provided by Penn State and the Computational and Information Systems Laboratory’s Cheyenne supercomputer provided by NCAR. This research was partially funded by the Penn State College of Earth and Mineral Sciences, a Penn State Institutes of Energy and the Environment seed grant, and the National Science Foundation (NSF) under Grant No. ICER-2019758. The participation of S.J.G. was partially supported by NSF CAIG grant No. 2425658. This material is based upon ongoing work at NCAR, which is a major facility sponsored by the NSF under Cooperative Agreement No. 1852977. Author contributions are as follows: conceptualization: D.F., D.J.G., S.J.G., and J.S.; methodology: D.F., D.J.G., C.S., S.J.G., and J.S.; investigation: all authors; supervision: S.J.G. and D.J.G.; writing—original draft: D.F.; and writing—review and editing: all authors. 

%
%
\datastatement
Processed training and testing data are available online at \url{https://doi.org/10.26208/6Y59-0R80}

%
\newpage

\appendix

%



%




\bibliographystyle{ametsocV6}
\bibliography{references}

\end{document}